\documentclass[english,prd,superscriptaddress,nofootinbib,preprintnumbers,twocolumn,showpacs]{revtex4}
\usepackage[utf8]{inputenc}
\usepackage[english]{babel}
\usepackage{amsmath}
\usepackage{amsfonts}
\usepackage{amssymb}
\usepackage{epsfig}
\usepackage{graphics,psfrag,rotating}
\usepackage{graphicx}
\usepackage{dcolumn}
\usepackage{bm}
\usepackage{textcomp}
\usepackage{keyval}
\usepackage{epstopdf}
\usepackage{color}
\usepackage[usenames,dvipsnames,svgnames]{xcolor}
\usepackage[T1]{fontenc}
\usepackage{multirow}
\usepackage{float}
\usepackage{subfigure}
\usepackage{xcolor}

\begin{document}
\title{Constraining spatial variations of the fine structure constant using clusters of galaxies and  {\it Planck} data}
\author{I. de Martino}
\affiliation{Department of Theoretical Physics and History of Science, 
University of the Basque Country UPV/EHU, Faculty of Science
and Technology, Barrio Sarriena s/n, 48940 Leioa, Spain.}
\email[]{ivan.demartino1983@gmail.com}
\author{C.J.A.P. Martins}
\affiliation{Centro de Astrofisica da Universidade do Porto, Rua das Estrelas s/n, 4150-762 Porto, Portugal}
\affiliation{Instituto de Astrof\'{\i}sica e Ci\^encias do Espa\c co, CAUP, Rua das Estrelas, 4150-762 Porto, Portugal.}
\email[]{Carlos.Martins@astro.up.pt}
\author{H. Ebeling}
\affiliation{Institute for Astronomy, University of Hawaii, Honolulu, HI 96822 USA.}
\email[]{ebeling@ifa.hawaii.edu}
\author{D. Kocevski}
\affiliation{Department of Physics and Astronomy, Colby College, Waterville, ME 04901, USA.}
\email[]{dale.kocevski@colby.edu}


\begin{abstract}
We propose an improved methodology to constrain spatial variations of the fine structure constant 
using clusters of galaxies. We use the {\it Planck} 2013 data to measure the thermal Sunyaev-Zeldovich effect 
at the location of 618 X-ray selected clusters. We then use a Monte Carlo Markov Chain algorithm to obtain the 
temperature of the Cosmic Microwave Background at the location of each galaxy cluster. When fitting three different 
phenomenological parameterizations allowing for monopole and dipole amplitudes in the value of the fine structure 
constant we improve the results of earlier analysis involving clusters and the CMB power spectrum, and we also found 
that the best-fit direction of a hypothetical dipole is compatible with the direction of other known anomalies. 
Although the constraining power of our current datasets do not allow us to test the indications of a fine-structure constant dipole 
obtained though high-resolution optical/UV spectroscopy, our results do highlight that clusters of galaxies will 
be a very powerful tool to probe fundamental physics at low redshift.
\end{abstract}

\maketitle

\section{Introduction}\label{sec:intro}

Recent observational studies suggest that the fine structure constant could vary over the sky.
Such spatial variation has been mainly studied using a large archival dataset of metal absorption lines in the
redshift range $z=[0.3 - 4.2]$ along the line-of-sight of bright quasars, and
its dipole amplitude has been measured to be $\Delta\alpha /\alpha =(12\pm 2)\times 10^{-6}$  
with a best fit dipole direction  (RA,DEC)=$(17.3\pm1 hours, -61^\circ \pm 10^\circ)$ 
\cite{webb2011, King2012}. Other analyses have confirmed these results 
\cite{Berengut2011, Berengut2012}, although the inclusion of more recent measurements reduces the allowed
amplitude by about twenty percent \cite{Pinho2016}, to a maximum of about 8 parts per million (with no significant
changes to the preferred direction). Although it is clear that some systematic effects are present in the archival
data \cite{whitmore}, it is presently unclear if they could explain the above results \cite{Cameron2012, Cameron2013}.

It has also been claimed that the best-fit dipole direction is comparable to the direction of other so-called
anomalies and/or dipoles observed in the CMB, in supernova and in bulk flows data  \cite{Mariano2012, Mariano2013}. On
the other hand, it is several degrees far away  from the directions of the intrinsic CMB and the Dark Flow dipoles, 
from the kinetic CMB asymmetry, and from the CMB cold spot location \cite{Kogut1993, darkflow2008, darkflow2010, darkflow2011, darkflow2012, darkflow2015, 
Vielva2004, planck13_XXIII, Zhao2012, Zhao2013, Zhao2014, Zhao2015, Zhao2016}.
Studies that specifically look for a dipole modulation of the $\alpha$ anisotropies on the 
Cosmic Microwave Background (CMB) power spectrum do not report any detection \cite{planck_int_24, bryan2015}, although their current sensitivity is
about one thousand times worse than the amplitude of the dipole inferred from high-resolution spectroscopy.
In these circumstances it is important to explore additional independent tests that may verify or rule out the spectroscopic result. In the present
work we focus on a complementary analysis, at lower redshifts, which can be carried out using current and 
forthcoming multi-frequency measurements of galaxy clusters. 

Clusters of galaxies contain a hot Intra-Cluster Medium (ICM) which reaches 
temperatures in the range $T_{\rm e}\sim 1-10$ keV. The hot electrons of the ICM
lose energy via inverse Compton scattering with the CMB photons. This process
produces secondary anisotropies on the CMB power spectrum, which
have two components: the thermal Sunyaev Zeldovich effect (TSZ, \cite{tsz}) due to the thermal
motion of the electron in ICM medium, and the kinematic one (KSZ, \cite{ksz}) due to the 
peculiar velocity of the cluster with respect to the isotropic CMB frame.
The TSZ anisotropies induced by clusters of galaxies along the line of sight $l$ are
usually expressed in terms of the Comptonization parameter $Y_c$
\begin{equation}\label{eq:TSZ}
\Delta T_{TSZ}=T_0G(x)Y_c= T_0G(x)\frac{k_B\sigma_{\rm T}}{mc^2}\int_l T_{\rm e}(l)n_{\rm e}(l)dl, 
\end{equation}
where $G(\tilde{\nu})$ is the TSZ spectral dependence, $\sigma_T$ is the Thomson cross section,  
$m_e$ is the electron mass, $c$ is the speed of light and $k_B$ is the Boltzmann constant, and
$n_e(l)$ and $T_e(l)$ are the electron density and temperature along the line of sight. 
$T_0$ is the current value of the CMB black-body temperature $T_0=2.725\pm 0.002$K \cite{fixsen}, 
and in the non relativistic limit ($T_e \approx$ few keV) the spectral dependence
has the following functional form: $G(\tilde{\nu})= \tilde{\nu}{\rm coth}(\tilde{\nu}/2)-4$, 
where $\tilde{\nu}=h\nu(z)/k_BT_{CMB}(z)$ is the reduced frequency,
$h$ is the Planck constant and $T_{CMB}(z)$ is the CMB black-body temperature at the cluster location.

In the concordance $\Lambda$CDM model, the standard evolution of the CMB black-body temperature is 
$T_{CMB, std}(z)=T_0(1+z)$. However, in particular classes of models, the evolution of the CMB temperature 
can be related to that of other observables. For example in models where an evolving scalar field is coupled 
to the Maxwell $F^2$ term in the matter Lagrangian, photons can be converted into scalar particles violating 
the photon number conservation. Thus, there will be both variations of the fine-structure  constant and violations of
the standard $T_{CMB}(z)$ law. 
{An example of this is Bekenstein-Sanvik-Barrow-Magueijo (BSBM) class of models \cite{BSBM}, where the electric charge is allowed to vary. 
Although such theories preserve the local gauge and Lorentz invariance, the fine structure constant will vary during the matter dominated 
era. The corresponding action is}
\begin{equation}
 S=\int{d^4x \sqrt{-g}(\mathcal{L}_{g}+\mathcal{L}_m-\frac{\omega}{2}\partial_\mu\phi\partial^\mu\phi-e^{-2\phi}\mathcal{L}_{em})},
\end{equation}
{ where $\mathcal{L}_{g}+\mathcal{L}_m$ is the standard Hilbert-Einstein Lagrangian plus the matter fields, the third term is the kinetic term for the scalar field $\phi$ and, finally, the last term couples the scalar field with the standard
electromagnetic Lagrangian $\mathcal{L}_{em}=\frac{F^{\mu\nu}F_{\mu\nu}}{4}$. In this case equation governing the evolution of the radiation energy reads}
\begin{equation}
 \dot{\rho}_\gamma+4H\rho_\gamma=2\dot{\phi}\rho_\gamma
\end{equation}
{which, assuming the adiabaticity, leads to} \cite{Avgoustidis2014}
\begin{equation}\label{eq:tcmb}
\frac{T_{CMB}(z)}{T_0}\sim(1+z)\left(1+\epsilon\frac{\Delta\alpha}{\alpha}\right)\,,
\end{equation}
or alternatively
\begin{equation}\label{eq:tcmb2}
\frac{\Delta T_{CMB}}{T}=\frac{T_{CMB}(z)-T_{CMB, std}(z)}{T_{CMB, std}(z)}\sim\epsilon\frac{\Delta\alpha}{\alpha}\,.
\end{equation}
The coefficient $\epsilon$ depends on the specific model being assumed, but it is generically expected to be of 
order unity. In particular, if one assumes the somewhat simplistic adiabatic limit, then one can show that
 $\epsilon=1/4$  \cite{Avgoustidis2014}. Therefore, if one is able to determine 
the CMB temperature at the cluster location using the SZ effect, the relation in eq. \eqref{eq:tcmb} or \eqref{eq:tcmb2}
can be used as a phenomenological relation to observationally test the spatial variation  of $\alpha$. 

Based on Planck 2013 Nominal data, the dipole variation of $\alpha$ has been constrained using 
the scaling relation between the SZ emission and its X-ray counterpart \cite{Galli2013}. With percent level sensitivity,
the dipole amplitude was found to be consistent with the standard evolution. However, the Planck SZ data was treated
fixing the evolution of the CMB temperature, and given the theoretical expectations discussed in the previous
paragraph this could potentially bias the final results.

In this work we will apply the techniques developed in \cite{demartino2015}
to measure the TSZ emission at the location of 618 X-ray selected clusters. Those
measurements will be used to test the spatial variation of the fine structure constant
in the redshift range $z=[0 - 0.3]$.  The outline
of the paper is the following: Section 2 will be devoted to describe the X-ray cluster catalog and illustrate the cleaning
procedure of the {\it Planck} 2013 Nominal maps; in Section 3 we will explain the methodology used to measure the CMB temperature
at cluster locations as well as to carry out the fitting procedure to test for spatial variations of $\alpha$; in
Section 4 we illustrate and discuss the results of our analysis; and finally, in Section 5, we give our main conclusions.

\section{Data.} \label{sec:data}

We use a sample of 618 X-ray clusters and the {\it Planck} 2013 Nominal maps\footnote{Data are available at 
http://www.cosmos.esa.int/web/planck} to determine the CMB temperature at each cluster's location on the sky.
Maps were originally released in a Healpix format with resolution $N_{side} = 2048$ \cite{gorski2005}. 

\subsection{X-ray Cluster Catalog.}\label{sec:xray_catalog}

The clusters of galaxies were selected from three ROSAT X-ray flux limited surveys: the 
ROSAT-ESO Flux Limited X-ray catalog (REFLEX, \cite{bohringer04}), 
the extended Brightest Cluster Sample (eBCS, \cite{ebeling98, ebeling00})
and the Clusters in the Zone of Avoidance (CIZA, \cite{ebeling02}). A
deeper discussion about the methodology used to compile the catalog is given
in \cite{kocevski-ebeling06}. The final catalog contains 782 clusters 
with well measured positions, spectroscopic redshifts,
X-ray fluxes in the [0.1-2.4]keV band and angular extents of the region 
emitting 99\% of the X-ray flux. It also lists the X-ray temperature
derived from the $L_X-T_X$ relation of \cite{white} and the core radii $r_c$,
central electron density $n_{e,0}$ and the central Comptonization parameter $y_{c,0}$
obtained by fitting a $\beta=2/3$ model to the ROSAT data.
Of those clusters, 618 are outside the  point-sources and the Planck galactic masks \cite{planck13_XXVIII, planck13_XXIX}.
Using the $r_{500}-L_X$ relation of \cite{bohringer07}, we
derived  $r_{500}$, the scale radius at which the mean overdensity of
the cluster is 500 times the critical density, and the corresponding angular 
size $\theta_{500}=r_{500}/d_A$. The angular diameter distance of each cluster, $d_A(z)$, is computed in 
the $\Lambda$CDM model assuming the best-fit parameters given in \cite{planck13_XVI}---this is a reasonable assumption
since in these models the scalar field is expected to have a negligible role in the underlying cosmological dynamics.
In addition, we defined $M_{500}$ as the mass enclosed within a sphere of radius $r_{500}$.

\begin{figure*}[!ht]
 \centering
 \includegraphics[width=1.2\columnwidth, angle=270]{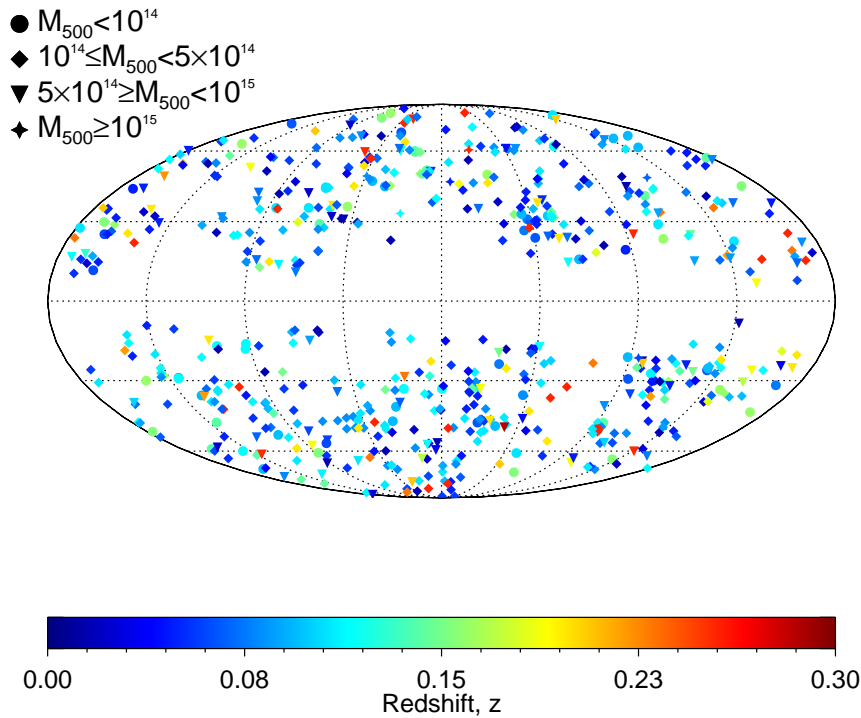}
\caption{Distribution on the sky of the clusters in our sample.}
\label{fig:clust_distrib}
\end{figure*}

The galaxy clusters in our sample have masses in the interval 
$M_{500}=[0.02-1.5]\times 10^{15}$M$_\odot$ and are located at 
redshifts $z\le 0.3$; thus our redshifts are relatively lower compared with other catalogs 
used for similar analysis, but our sample does contain almost 10 times more clusters 
homogeneously distributed over the sky.  Figure \ref{fig:clust_distrib} shows the spatial distribution on the sky
of the clusters used in this analysis. Different colors and symbols are used to illustrate how the clusters
are distributed in terms of redshift and mass.

For comparison, the catalog used by Galli (2013) \cite{Galli2013} to constraint the spatial variation
of the fine structure constant contains 61 SZ selected clusters \cite{planck_early_XI}. 
However, those clusters are identified in the CMB maps  using the standard temperature evolution $T(z)_{CMB, std}$.
Thus, the sample will contain galaxy clusters that mimic this behavior and could bias the final results. 
Therefore, it is extremely important to carry out tests using only X-ray selected clusters 
since they will not be affected by these biases.

{We also note (as demonstrated by the comparison of the performance of SPT and PLANCK) that  high 
angular resolution and precautions against correlated noise are crucial for SZ missions to be efficient and 
reliable at detecting clusters, a point stressed by the Planck Collaboration when they wrote "SZ candidates 
with no detection at all in the ROSAT All-Sky Survey are almost certainly false" \citep{planck13}. 
Existing X-ray selected samples of only the most massive clusters \citep[e.g.,][]{ebeling01,ebeling13} 
are thus ideally suited for work like ours and we will use such an extended X-ray cluster sample as soon 
as it becomes available to improve upon the analysis presented here.}

\subsection{Foreground cleaned {\it Planck} Nominal Maps.}\label{sec:cleaning}

In March 2013 the Planck Collaboration made publicly available its Nominal maps covering the frequency channels from
30 to 857 GHz. The maps contain intrinsic CMB, TSZ and KSZ emission, instrumental noise and foreground 
emissions due to galactic dust, CO lines, synchrotron radiation and point and extended infrared sources. 
Although the TSZ effect has a peculiar spectral dependence  that allows us to distinguish it from other components,
it can not be reliably detected without reducing the foreground emissions and the cosmological CMB signal. 

For that purpose, the Planck Collaboration has constructed  a full sky map of the TSZ effect
applying an internal linear combination technique to the 100 to 857 GHz frequency channel maps \cite{planck13_XXI}. 
When applying the procedure they fixed the spectral dependence $G(\tilde{\nu})$ to the standard one. 
Therefore, those data are not useful to estimate the CMB temperature at cluster location since
$T_{CMB}(z)$ has been fixed to the adiabatic evolution. Nevertheless, 
we apply a methodology developed in \cite{diego2002}  that does not make any of the typical assumptions
about the frequency dependence of the different astrophysical signals, nor the cluster profile. 
This methodology has been successfully applied to extract the TSZ signal 
from the Planck maps and to test the blackbody temperature-redshift relation, the SZ/X-Ray correlations, 
and also to constrain some models of modified gravity \cite{diego2003, demartino2015, Luzzi2015, demartino2016a, demartino2016b}.

{\it Before applying the cleaning procedure, we remove the intrinsic CMB monopole and dipole from the Nominal
maps}. Then, the procedure was applied only to the High Frequency Instrument (HFI) data,
covering the frequency range from 100 to 857 GHz, since they have better angular resolution 
$\theta_{FWHM}<10'$ and lower instrumental noise than the data from the 
Low Frequency Instrument (from 30 to 70 GHz). Let us briefly report the main steps  
of cleaning procedure developed in \cite{demartino2015}:

\textbullet \, we brought all channels to a common 10 arcmin angular resolution corresponding to the 
lowest frequency channel considered in this study ($100$ GHz); \\

\textbullet \, we removed the cosmological CMB and KSZ signals  subtracting the LGMCA 
CMB template \cite{bobin2013, bobin2014} at each {\it Planck} channel. It has been previously
demonstrated that this step does not introduce any distortion to the TSZ emission \cite{demartino2015};\\

\textbullet \, we used the CO Type 2 maps provided by the Planck Collaboration \cite{planck13_XIII}, 
to remove CO emission at 100 and 217~GHz;\\

\textbullet \, we excised point sources and foreground emission close to the
Galactic Plane applying the PCCS-SZ-Union mask \cite{planck13_XVIII, planck13_XXIX}; \\

\textbullet \, finally, we used the 857~GHz channel as a dust-template to reduce 
the thermal dust emission \cite{diego2002, planck13_XI, planck13_XII}. \\

At the end of the procedure, we have a
foreground cleaned patch $\mathcal{P}(\nu, x)$  centered on the position $x$ 
of each selected cluster in our catalog. From those patches, we have measured the average TSZ temperature 
fluctuations\footnote{Hereafter, for simplicity, we will avoid specifying the angular size over which we measure the signal,
being understood that the averaged quantities are always evaluated on discs of radius $\theta_{500}$, and indicated with
a bar.} ($\delta \bar{T}/T_0 (\nu)$), where the average was evaluated over discs 
with angular extent equal to $\theta_{500}$.

To compute the error bars, we carried out 1,000 random simulations 
evaluating the mean temperature fluctuations from patches randomly placed 
out of the cluster positions. Each patch was cleaned following the same procedure adopted for 
the real cluster population. To avoid overlapping with real X-ray clusters,
the patch centers were placed  at least 2 degrees away from the cluster centers.
After verifying that  the average over all simulations was consistent with zero, 
we computed the correlation matrix between different frequencies as  
\begin{equation}
C(\nu_i,\nu_j)=\frac{\langle [\delta \bar{T}(\nu_i)-\mu(\nu_i)]
[\delta  \bar{T}(\nu_j)-\mu(\nu_j)]\rangle}{\sigma(\nu_i)\sigma(\nu_j)} ,
\label{eq:Cij}
\end{equation}
where the average was computed over all simulations, $\mu(\nu_i)=\langle\delta  \bar{T}(\nu_i)\rangle$, and 
$\sigma(\nu_i)=\langle [\delta  \bar{T}(\nu_i) -\mu(\nu_i)]^2\rangle^{1/2}$.
The error bars on the TSZ temperature anisotropies of each cluster are the square root of 
the diagonal elements of the correlation matrix. 

\begin{figure*}[!ht]
 \centering
 \includegraphics[width=1.9\columnwidth]{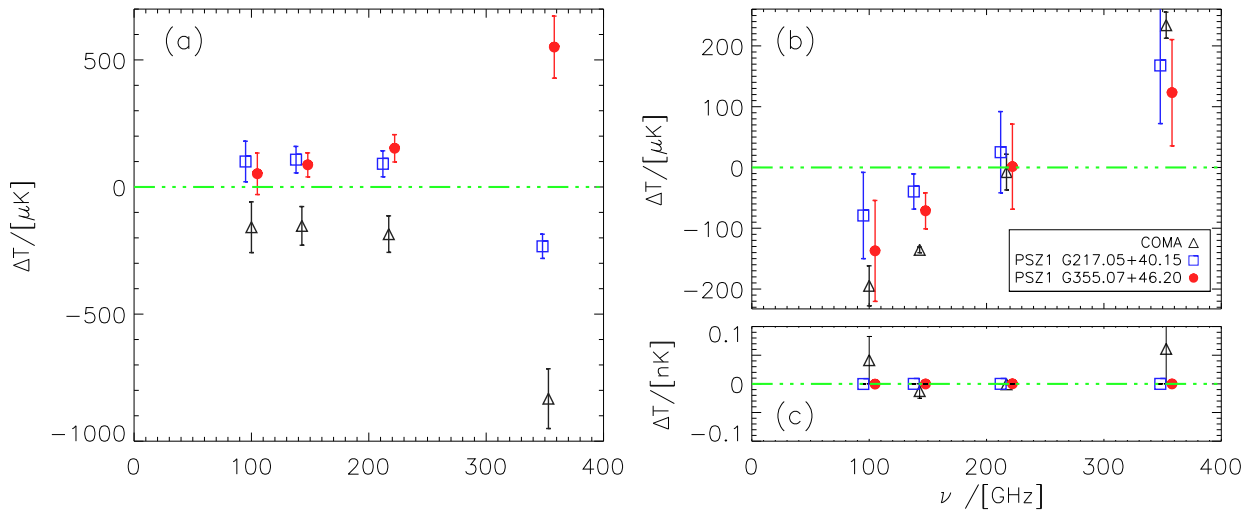}
 \vspace*{-4cm}
\caption{(a) Average of the temperature anisotropy as a function of frequency
for the three selected clusters  evaluated on the Planck 2013 Nominal map on discs of radius
$\theta_{500}$.   (b) TSZ anisotropy measured on a disc of size $\theta_{500}$ on our foreground cleaned patches. 
(c) Average temperature 
anisotropy evaluated in a ring placed around the three clusters. 
Red circles and blue squares have been shifted by $\pm5$ GHz for easier viewing.}
\label{fig2}
\end{figure*}

To illustrate the effectiveness of our cleaning procedure, in Fig.~\ref{fig2} we plot the 
average temperature anisotropy of three {\it Planck} clusters at the different HFI frequency channels.
We have chosen three {\it Planck} clusters covering the whole redshift range of our catalog. Specifically,
we plot: Coma (Abel 1656) which is located at $z=0.0231$, has a mass $M_{500}=5.3\times10^{14}M\odot$
and an angular extent $\theta_{500}\sim50'$; PSZ1 G217.05+40.15 at $z=0.1357$, with mass and angular extent
equal to $M_{500}=4.6\times10^{14}M\odot$ and $\theta_{500}\sim9'$, respectively; and, finally, 
PSZ1 G355.07+46.20 at $z=0.2153$, with  $M_{500}=1.46\times10^{15}M\odot$ and $\theta_{500}\sim9'$.
In Figure ~\ref{fig2}a we plot the averaged temperature anisotropies  measured on the Nominal
maps. This clearly shows that intrinsic CMB anisotropies, thermal dust and other foreground emissions dominate
over the TSZ effect. In Figure ~\ref{fig2}b we show that the mean TSZ temperature anisotropies 
of those clusters measured on the foreground cleaned patches  are preserved by the cleaning procedure.
In fact, they are negative at 100 and 143 GHz,  zero at 217 GHz (at $1\sigma$ level), and positive at 353 GHz.
We do not include the 545 GHz channel in our analysis since here our cleaning procedure no longer gives reliable 
results and the TSZ anisotropies are strongly contaminated by dust residuals. 
Finally, in Figure ~\ref{fig2}c, for the same clusters we report  the mean temperature anisotropies 
within the ring placed around each cluster and having the inner and outer 
radii equal to  $4$ and $5$ times $\theta_{500}$, respectively. This shows that the residuals 
around the clusters are negligible compared with the TSZ emission. We also checked the mean temperature anisotropy
around other clusters and we found it to be $\le 10^{-3}\mu$K. 

\section{Methodology}

To estimate the CMB temperature at the cluster location as well as to
constrain the spatial variations of the fine structure constant, we
explore the relevant parameter space corresponding using a Monte Carlo Markov Chain (MCMC) technique. 
We always run four independent chains employing  the Metropolis-Hastings sampling 
algorithm \cite{Metropolis1953, Hastings1970} with different (randomly set) starting
points. The chains stop when they contain at least 30,000 steps and satisfy
the Gelman-Rubin criteria \citep{Gelman1992}.  
The step size is adapted in order to reach an optimal acceptance rate between 20\% and 50\% \cite{Gelman1996, Roberts1997}.
Finally, the four chains are merged and the marginalized distributions are computed using all the steps.
The expectation value of the 1D marginalized likelihood distribution ($\mathcal{L}({\bf p}_i)$) and its variance 
were computed as \cite{Spergel2003}
\begin{align}
& \langle {p}_i\rangle=\int d^{N_s}{\bf p} \mathcal{L}({\bf p}) {p}_i,\label{eq:bestfit}\\
&  \sigma^2_i=\int d^{N_s}{\bf p}  \mathcal{L}({\bf p}) ({p}_i -  \langle{p}_i\rangle)^2\label{eq:errorbestfit},
\end{align}
where $\langle {p}_i\rangle$ denotes the expectation value of the parameter ${p}_i$, and $N_s$ is the dimension
of the parameter space explored by the MCMC.

\subsection{Obtaining $T_{CMB}(z)$ from the foreground cleaned {\it Planck} Nominal maps}\label{sec:Tcmb_MCMC}

We measure the averaged TSZ emission on cleaned patches around each cluster in our catalog (see Section \ref{sec:data}).
Then, from eq. \eqref{eq:TSZ}, we predict the theoretical averaged TSZ anisotropies at the same aperture 
\begin{equation}\label{eq:<TSZ>}
\Delta \bar{T}({\bf p}, \nu_i)/T_0 =G(\nu_i, T_{CMB}(z)) \bar{Y}_c,
\end{equation}
where ${\bf p}=[T_{CMB}(z),\bar{Y}_c]$ are the free parameters of the model, 
and $\bar{Y}_c$ is 
the averaged Comptonization parameter. In general, a model predicting the cluster profile 
should be assumed to correctly predict $\bar{Y}_c$.  
Since we are not interested  in studying the details of the cluster physics, we can adopt an alternative approach 
based on considering $\bar{Y}_c$ as a extra free parameter \cite{demartino2015}. 
Finally, we fit our theoretical predictions to the data computing the likelihood $-2\log{\cal L}=\chi^2({ \bf p})$ as
\begin{equation}
-2\log{\cal L}=\chi^2 ({ \bf p})=\Sigma_{i,j=0}^{N} 
\Delta \bar{T}_{i}({\bf p})C_{ij}^{-1}
\Delta \bar{T}_{j}({\bf p}),
\label{eq:chi}
\end{equation}
where $\Delta \bar{T}_{i}({\bf p})\equiv\dfrac{\Delta \bar{T}({\bf p}, 
\nu_i)}{T_0}-\dfrac{\delta \bar{T}(\nu_i)}{T_0}$,
$N=4$ is the number of data points (the four frequencies), and 
$C_{ij}$ is the correlation matrix given in eq. \eqref{eq:Cij}.
When computing the likelihood we neglect the error on the CMB blackbody  
temperature $T_0$ since it is negligible with respect to the error 
on the temperature anisotropies. To improve the description of the TSZ effect,
we also include relativistic corrections in the electron temperature 
up to fourth order \cite{itoh1998, nozawa1998, nozawa2006}.

The parameter space explored by our MCMC is chosen on physical grounds: for the $T_{CMB}(z)$ parameter we
follow the prescription given in \cite{Luzzi2015}. Thus, we allow for a broad Gaussian prior centered on the standard value
of the CMB temperature ($T_0(1+z)$) with a standard deviation of $0.5\rm{K}(1+z)$. This choice avoids
the possibility that MCMC chains spend a lot of time exploring unphysical regions of the parameter space. 

For the $\bar{Y}_c $, we adopt the following procedure: as mentioned in Sect. \ref{sec:data}
our catalog lists $y_{c,0}$ obtained by fitting  a $\beta(=2/3$)-model to the X-ray data. Thus, 
we used $y_{c,0}$ to predict the Compton emission at $\theta_{500}$ for all clusters assuming a $\beta(=2/3$)-model 
for the cluster profile.  Since this model correctly predicts the TSZ anisotropies 
only within the X-ray emitting region \cite{atrio2008} while it overestimates them at $\theta_{500}$, 
we can not use  $\bar{Y}_{c, \beta=2/3}$ to reduce the number of free parameters. 
Such averaged Comptonization parameter, $\bar{Y}_{c, \beta=2/3}$, will be used as reference 
value to choose the prior for each individual galaxy cluster in our catalog.
Given that for our cluster catalog  $\bar{Y}_{c, \beta=2/3}/\bar{Y}_c  \sim 1.1$ (on average) \cite{demartino2016a},
we conservatively choose a flat prior allowing for a 50\% variation around 
the reference values.

We checked that relaxing the above assumptions and using instead a flat prior 
on both $T_{CMB}(z)$ and the Comptonization parameter does not change  the final accuracy of our estimation.
We set very broad, but fixed, priors for all clusters allowing the CMB temperature and 
Comptonization parameter to vary between $[0 - 5]$K and $[0 - 10^{-3}]$, respectively.  These  choices have the only effect 
of increasing the computing time.
 
\begin{figure}[!ht]
 \centering
 \includegraphics[width=1.0\columnwidth]{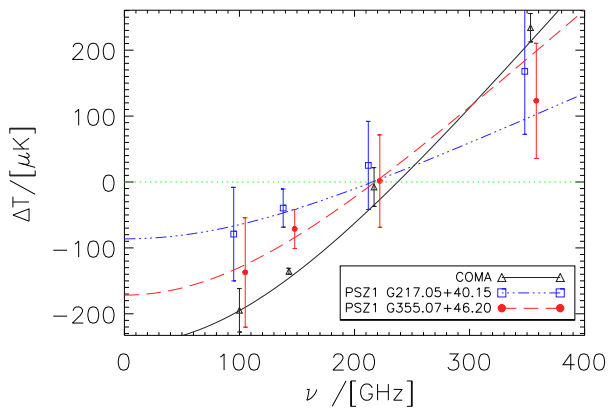}
\caption{Best fit to the TSZ temperature anisotropies of the three clusters represented in Figure \ref{fig2}. 
For each cluster we plot the best fit model obtained considering the relativistic corrections to
the frequency dependence.}
\label{fig4}
\end{figure}

In Figure \ref{fig4}, we show the effectiveness of our procedure in extracting the CMB temperature at the cluster location.
We plot the average TSZ temperature anisotropies of the three clusters represented in Figure \ref{fig2}
and their errors, for the different frequencies. The  lines correspond to the best fit model,
$\bar{Y}_c^* G(\nu, T_{CMB}^*(z))$, with  relativistic corrections.
Specifically, for Coma, PSZ1 G217.05+40.15 and PSZ1 G355.07+46.20 we found 
$\bar{Y}_c^*=[4.52\pm0.01, 1.6\pm0.3, 3.15\pm0.3]\times10^{-5}$ 
and $T_{CMB}^*(z)=[2.92\pm0.17, 3.07\pm0.19, 3.32\pm0.21]$K, respectively.
We have checked the effect on the estimation of the CMB temperature of the inclusion (or absence) of relativistic
corrections and found it to always be less than 3\% even for the three clusters represented in Figure \ref{fig4}
having the X-ray temperature $>6$ keV. Finally, the $\chi^2$ per degree of freedom
of the three best fit models is  $\chi^2_{dof}=[0.96, 0.94, 0.89]$. 
These results further support our procedure. For the full sample, we checked that the relativistic corrections change less than 
$\sim1.8\%$ (on average) the estimation of the CMB temperature, while the $\chi^2_{dof}$ is within the range $[0.85,1.06]$.

\subsection{Model testing}

Using the measurements of the $T_{CMB}(z)$ we can estimate the 
$\alpha$-anisotropies at the location of our 618 X-ray selected clusters of galaxies as
\begin{equation}
 \frac{\Delta\alpha}{\alpha}=\epsilon^{-1}\left(1-\frac{T_{CMB}(z)}{T_0(1+z)}\right).
\end{equation}
Then, we can fit 3 different models to the 
CMB temperature data allowing for a dipole variation. Specifically, we consider the same models
used in \cite{webb2011} to fit quasar data and in \cite{Galli2013} to fit cluster data.
In such a way we are able to compare our results with the ones from other groups. As an additional
check we will also do the fit for the temperature measurements themselves (without expressing
them as $\alpha$ measurements). Thus, the models are: \\

{\bf Model 1.} a monopole plus dipole model of the $\Delta \alpha/\alpha$ measurements with 
the following functional form
\begin{equation}\label{eq:model1}
 \frac{\Delta\alpha}{\alpha} = m + d \cos(\Theta),
\end{equation}
where  $m$ is a monopole amplitude that allows for an offset due to the Earth motion, $d$ is the 
dipole amplitude, and $\Theta$ is the angle on the sky between the line of sight of each cluster 
and the best fit dipole direction.

{\bf Model 2.} Although our measurements are located at low redshift ($z<0.3$), we also consider a
generalization of the previous model where the look-back time, $r(z)=\int\frac{dz'}{H(z')}$, is introduced:
\begin{equation}\label{eq:model2}
 \frac{\Delta\alpha}{\alpha} = m + d r(z)\cos(\Theta).
\end{equation}
Here $r(z)$ is computed in the framework of the concordance $\Lambda$CDM model with the best 
fit parameters given in \cite{planck13_XVI}. This is a simple proxy to ascertain how sensitive the
data is to a possible redshift evolution. In this model, the dipole amplitude it is measured in units of $GLyr^{-1}$.

{\bf Model 3.} Finally, we also try to fit  a monopole plus dipole model  to the CMB temperature measurements. 
This means to fit the relation
\begin{equation}\label{eq:model3}
\frac{\Delta T}{T_0(1+z)} = m + d \cos(\Theta).
\end{equation} 
This is a very useful test to explore the possibility of an intrinsic dipole. In the class of models we are
considering, if there is a parts-per-million dipole in the values of the fine-structure constant, $\alpha$, there should
also be an additional CMB temperature dipole (that is, in addition to the standard one due to our motion) in the same
direction of the $\alpha$ dipole, and with $\mu$Kelvin amplitude. With a sample of hundreds of clusters we can therefore
constrain such hypothetical dipole variations.

For each model we carry out four different MCMC analyses: (A) we fix the monopole amplitude to zero for the Model 1 and 2, and
to unity for Model 3. The direction of the dipole is fixed to the best-fit value from Webb {\it et al.} \cite{webb2011}. The
model has one free parameter (i.e. the dipole amplitude);  (B) we still keep the direction of the dipole fixed at the best-fit
one from the spectroscopic measurements, but the monopole and dipole amplitudes are both free to vary. In (C) and (D) we repeat
the analysis as they are in (A) and (B) also allowing the direction of the dipole to vary. The different configurations of the 
parameter space corresponding to each analysis are summarized in Table \ref{tab:priors}.

\begin{table}
\begin{center}
\begin{tabular}{|c|ccccc|}
\hline
 {\bf Analysis} &  $m$ & $d$ & $RA$ & $DEC$ & $N_{par}$\\
 &  &   & ($^\circ$) & ($^\circ$) & \\
 \hline
(A) & $0$ or $1$ & $[-1,1]$  &      261.0      &  $-58.0$       & 1\\
(B) & $[-1,1]$   & $[-1,1]$  &      261.0      &  $-58.0$       & 2\\
(C) & $0$ or $1$ & $[-1,1]$  &      $[0, 360]$ &  $[-90, +90]$  & 3\\
(D) &  $[-1,1]$  & $[-1,1]$  &      $[0, 360]$ &  $[-90, +90]$  & 4\\
 \hline
\end{tabular}
\caption{Priors on the parameter space explored by the MCMC algorithm in each analysis.}\label{tab:priors}
\end{center}
\end{table} 

\section{Results and discussion}
\begin{table*}
\begin{center}
\begin{tabular}{|c|c|cccc|}
\hline
 &  &    $m$ & $d$ & $RA$ & $DEC$ \\
 & &   &   & ($^\circ$) & ($^\circ$) \\
 \hline
\parbox[t]{2mm}{\multirow{4}{*}{\rotatebox[origin=c]{90}{\text{Model 1} }}} & (A) &  0.0               & $-0.002\pm0.008$   &      261.0         &  $-58.0$   \\
                        & (B) &  $0.006\pm0.004$   & $-0.008\pm0.009$   &      261.0         &  $-58.0$        \\
                        & (C) &     0.0              & $-0.030\pm0.020$    &     $255.1\pm3.8$  &  $-63.2\pm2.6$  \\
                        & (D) &  $0.021\pm0.029$   & $-0.030\pm0.014$    &     $255.9\pm4.2$ &  $-55.3\pm5.8$   \\
 \hline
  \hline
\parbox[t]{2mm}{\multirow{4}{*}{\rotatebox[origin=c]{90}{\text{Model 2} }}} & (A) &   0.0          & $-0.003\pm0.003$ GLyr$^{-1}$   &      261.0          &  $-58.0$     \\
                        & (B) &  $0.006\pm0.005$   & $-0.003\pm0.005$ GLyr$^{-1}$ &      261.0          &  $-58.0$       \\
                        & (C) &      0.0             & $-0.042\pm0.049$ GLyr$^{-1}$   &      $261.6\pm16.1$ &  $-61.3\pm2.7$   \\
                        & (D) &  $0.019\pm0.011$   & $-0.027\pm0.051$ GLyr$^{-1}$   &      $245.0\pm12.9$ &  $-56.0\pm3.8$   \\ 
\hline
\hline
 \parbox[t]{2mm}{\multirow{4}{*}{\rotatebox[origin=c]{90}{\text{Model 3} }}} & (A) &  1.0                 & $-0.010\pm0.008$     &      261.0         &  $-58.0$  \\
                        & (B) &  $1.001\pm0.002$   & $-0.003\pm0.002$    &      261.0         &  $-58.0$       \\
                        & (C) &     1.0            & $-0.020\pm0.015$     &      $258.0\pm1.2$ &  $-64.0\pm1.1$  \\
                        & (D) &  $1.000\pm0.0001$  & $-0.018\pm0.015$    &      $258.4\pm1.9$ &  $-64.3\pm2.6$   \\
 \hline
\end{tabular}
\caption{Results of the analysis for the Models 1,2 and 3.}\label{tab:results}
\end{center}
\end{table*}

We fitted to our dataset the two models allowing for the spatial variation of the fine structure constant
(Model 1 and 2), where $\epsilon=1/4$ is fixed to its adiabatic value,
as well as the model to test the intrinsic dipole of the CMB temperature data (Model 3). To 
test those models we have used  {\it Planck} 2013 foreground cleaned Nominal maps and
a proprietary X-ray selected cluster catalog. To succinctly describe our pipeline, we have determined 
the CMB temperature at the location of each galaxy cluster in our catalog. Then, using 
eq. \eqref{eq:tcmb} or \eqref{eq:tcmb2} we are able to test the models in eqs. 
\eqref{eq:model1}, \eqref{eq:model2} and  \eqref{eq:model3}. For each model we carried out 
four different analyses corresponding to different configurations of the parameter space 
(see Table \ref{tab:priors}). The results of the Model 1, 2 and 3 are summarized in the Table \ref{tab:results}. 

In all cases, the dipole  and  the monopole amplitudes are
compatible with standard expectations at not more than 2$\sigma$. Our best constraints are obtained when the dipole direction
is set to the best fit one from Webb {\it et al.}  \cite{webb2011}. Specifically,  we obtained: for Model 1 (A) $d=-0.002\pm0.008$,
and (B) $m=0.006\pm0.004$  and $d=-0.008\pm0.009$; while for Model 2 (A) $-0.003\pm0.003$ GLyr$^{-1}$,
and (B) $m=0.0061\pm0.0045$  and $-0.003\pm0.005$ GLyr$^{-1}$.  Since our dataset is located at $z<0.3$, introducing
the dependence from the look-back time makes no noticeable difference (as expected) in the final results that are still compatible 
with zero.

\begin{figure*}[!ht]
 \centering
 \includegraphics[width=1.8\columnwidth]{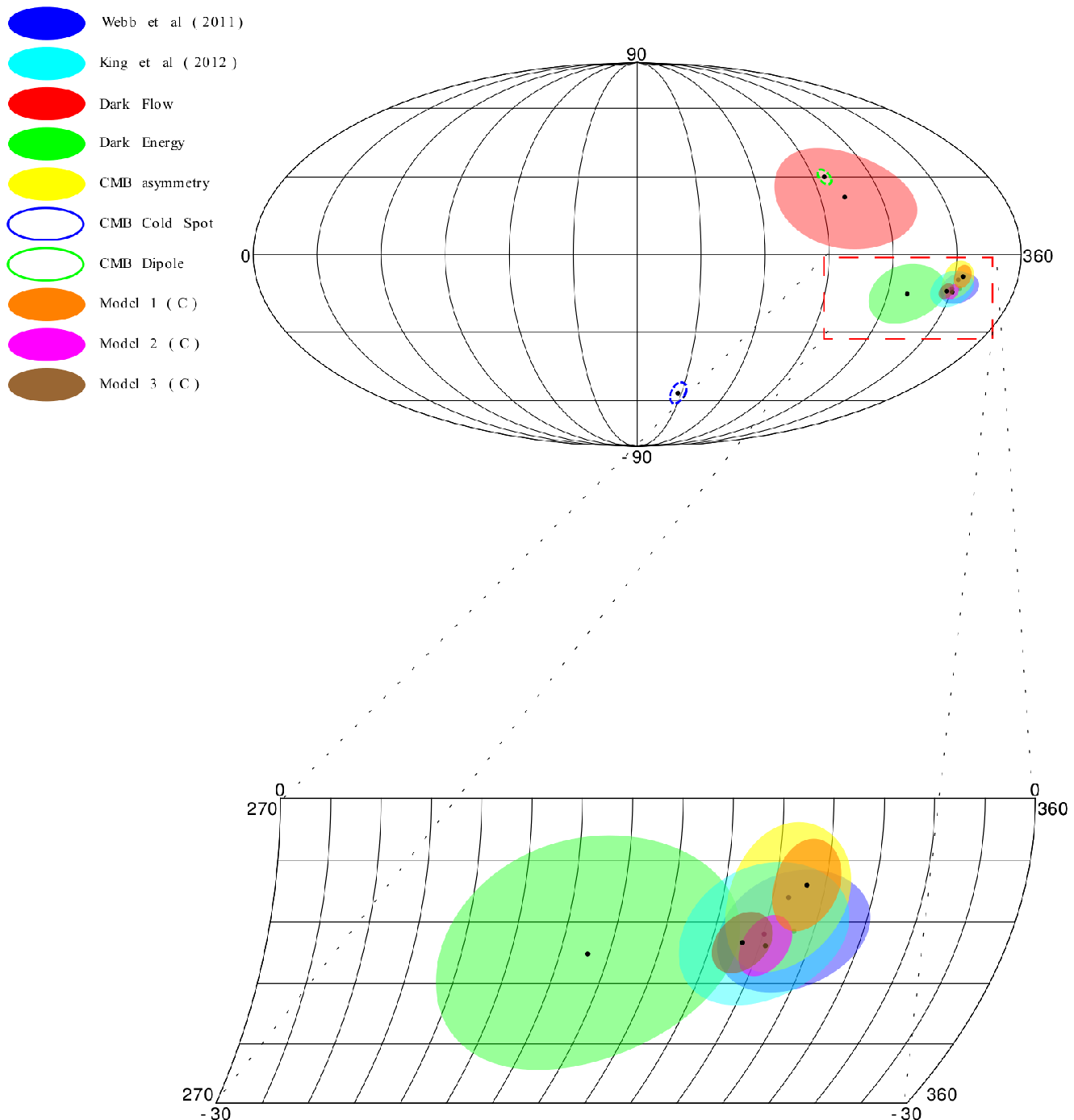}
\caption{Directions in galactic coordinates for the spatial variation of the fine structure constant 
 from \cite{webb2011} and \cite{King2012} in blue and cyan, respectively, for the Dark Energy dipoles from 
 \cite{Mariano2012} in green, Dark Flow direction from \cite{darkflow2008, darkflow2010, darkflow2011, darkflow2012, darkflow2015} in red,
for the CMB asymmetry  from {\cite{Mariano2013}} in yellow, and for our results from the analysis (C) of models 1,2 and 3 in
orange, magenta and brown, respectively. Finally, we also indicate in blue and green dashed circles the direction of the 
Cold Spot anomaly and the intrinsic CMB dipole, respectively \cite{Vielva2004, planck13_XXIII,Kogut1993}.}\label{fig5}
\end{figure*}

The constraints are significantly degraded when we also vary the direction of the best fit dipole. Nevertheless, we can still
get useful information. Specifically, the best fit directions of the three models are always compatible with each other,
and with the one from QSO data \cite{webb2011, King2012} and other CMB anomalies 
\cite{Mariano2012, Mariano2013} at $95\%$ CL, while the angular distance between our dipole direction and the directions of
the Dark Flow \cite{darkflow2008, darkflow2010,darkflow2011, darkflow2012, darkflow2015}, of the 
 CMB dipole direction \cite{Kogut1993} and of the Cold Spot anomaly\cite{ Vielva2004, planck13_XXIII} is
$\sim61^\circ$, $\sim70^\circ$, and $\sim100^\circ$, respectively, as shown in Figure \ref{fig5}. 
For an easier comparison we have also summarized the directions of the plotted anomalies in Table \ref{tab:anomalies}.

\begin{table}
\begin{center}
\begin{tabular}{|cccc|}
\hline
{\bf Anomalies} & $l$ & $b$ & {\bf Ref.} \\
 & ($^\circ$) & ($^\circ$) & \\
 \hline
 Cold Spot   & $207.8\pm5$ & $-56.3\pm5$ &\cite{Vielva2004, planck13_XXIII}\\
 CMB Dipole  & $276\pm3$ & $30\pm3$ &\cite{Kogut1993}\\
 Dark Flow   & $282\pm34$& $22\pm20$ &\cite{darkflow2008, darkflow2010, darkflow2011, darkflow2012, darkflow2015}\\
 Dark Energy & $309.4\pm18.0$&$-15.1\pm11.5$ &\cite{Mariano2012}\\
 CMB asymmetry & $331.9\pm7.3$& $-9.6\pm7.0$&\cite{Mariano2013}\\
 $\alpha$-dipole &$33.6\pm9.0$  &$-12.9\pm6.0$ & \cite{webb2011}\\
 $\alpha$-dipole & $330.1\pm10.0$ & $-13.16\pm7.0$& \cite{King2012}\\
 Model 1 (C) & $333.8\pm4.5$ & $-8.4\pm4.0$ & This work\\
 Model 2 (C) & $330.7\pm3.0$ & $-14.3\pm3.0$ &This work\\
 Model 3 (C) & $327.8\pm3.0$ & $-14.0\pm3.5$ &This work\\
 \hline    
\end{tabular} 
\caption{Galactic latitudes and longitudes of some known anomalies studied in literature, compared to the results of our
analysis (case C, for each of the 3 models we considered). The anomalies correspond
to the ones plotted in Figure \ref{fig5}.}\label{tab:anomalies}
\end{center}
\end{table}

\begin{figure}[!ht]
 \centering
 \includegraphics[width=0.55\columnwidth,angle=270]{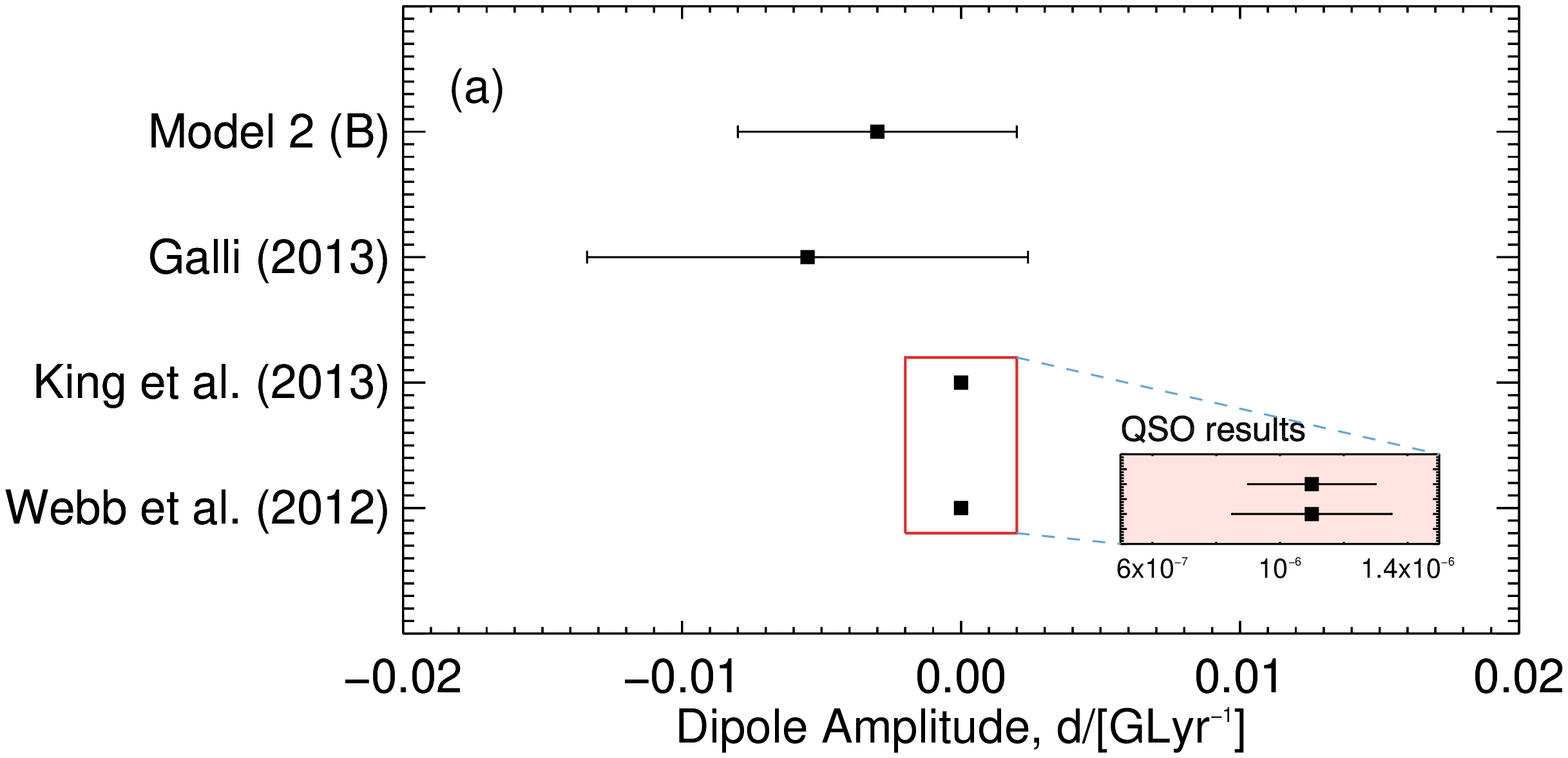}
 \includegraphics[width=0.55\columnwidth,angle=270]{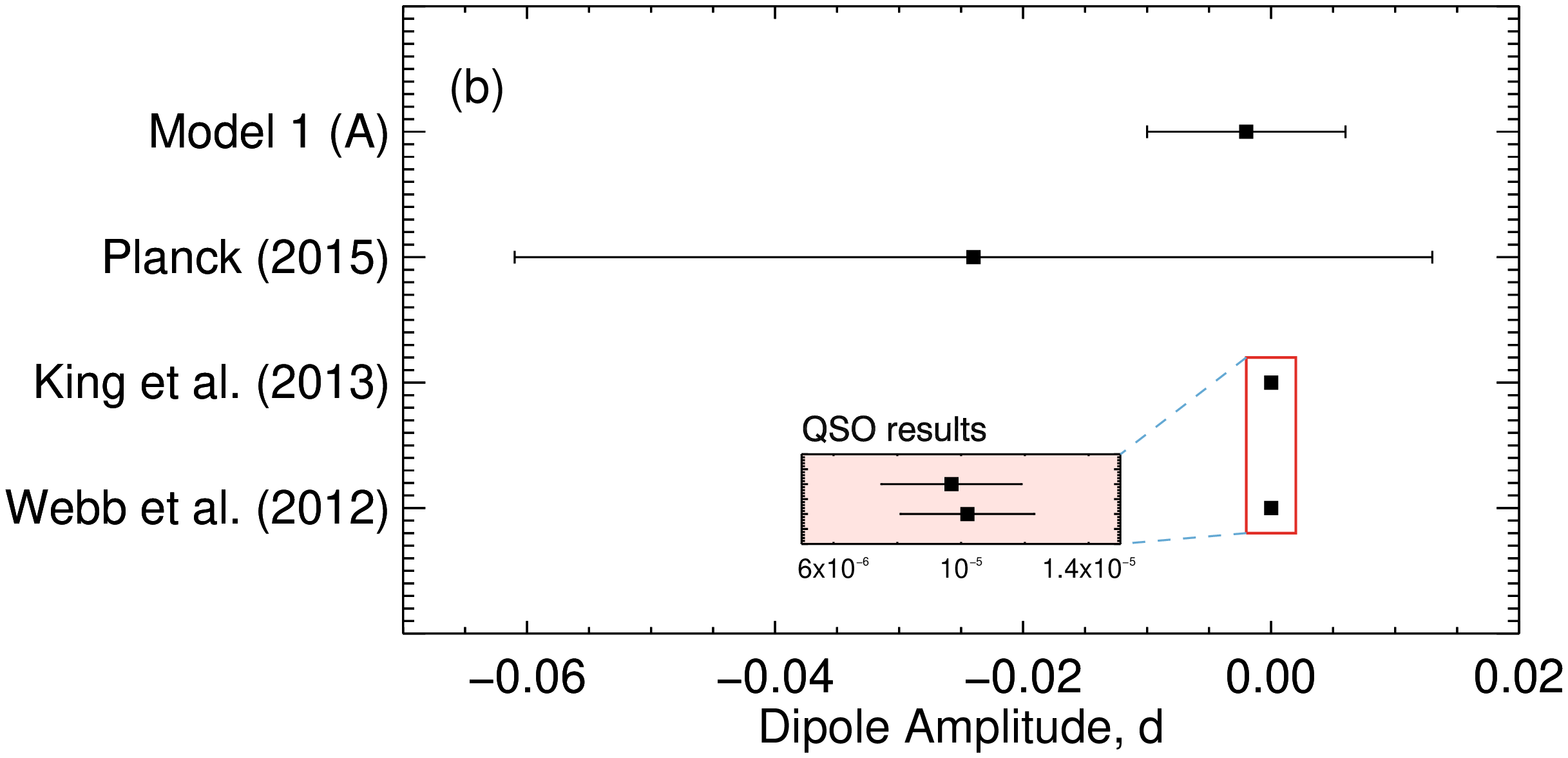}
\caption{Comparison of our constraint with the ones reported in previous works. See the main text for a discussion.}
\label{fig6}
\end{figure}

In Figure \ref{fig6}, we show the comparison our results from Model 1 and 2 with the results by other groups. 
Let us remark that as expected we are not competitive with the constraint from  \cite{webb2011} that 
gave $d=[(1.02\pm0.21)\times10^{-5};(1.1\pm0.25)\rm{GLyr}^{-1}]\times10^{-6}$ for the Model 1 and 2 respectively.
(See also \cite{Pinho2016} for a more recent analysis with additional spectroscopic data.) However, our results 
are important since they test for spatial variations of the fine structure constant in a different redshift 
range and with a different observable, thus having a different exposure to possible systematics.
 
Moreover, the constraints on the amplitude of dipole which we have obtained from our analysis in Model 2 improve by a factor
$\sim2.5$ (while being compatible with) the results obtained using the SZ selected cluster sample in \cite{Galli2013}, 
which was $d=(-0.005\pm0.0079)\rm{GLyr}^{-1}$. Last but not least, we also improve by a factor $\sim10$ the previous
constraint from the CMB analysis of Planck Collaboration and other groups \cite{planck_int_24, bryan2015}.

Finally, we briefly comment on the impact of our choice of $\epsilon=1/4$. This value stems from an assumption of
adiabaticity, which is further discussed in \cite{Avgoustidis2014} (and the value is exact in this limit). However, for
specific choices of models, its value may be different. For example, in the phenomenological class of models
described in \cite{hees} one has the approximate value $\epsilon\sim0.12$, and one may therefore ask how would our
results be affected by this change. We have tested this, specifically for the case of Model 2, and in all the
four analyses (A-D) we found that at the one sigma level the results were compatible with the ones we have
reported for $\epsilon=1/4$. Thus current data is unable to distinguish between different values of $\epsilon$
(provided that they are of order unity, which is expected to be the case).

\section{Conclusions}
The standard Hot Big Bang model predicts that the Universe evolves adiabatically. Thus, the CMB
temperature-redshift relation is linear. However more complicated prescriptions, including models
in which the photon number conservation is violated, will change the way in which the CMB
temperature scales with the redshift. We have considered models in which a scalar field is coupled
to the Maxwell term in the matter Lagrangian giving rise to the variation of the fine structure
constant together with the variation of the CMB temperature, as in Eqs. \eqref{eq:tcmb} or 
\eqref{eq:tcmb2}, and used cluster and Planck data to constrain them.

We started our analysis by cleaning the Planck 2013 Nominal maps and extracted the CMB temperature at
the location of 618 X-ray selected clusters using the TSZ multi-frequency measurements.
We then estimated the values of $\alpha$ at the location of each of the clusters
in our sample using the eq. \eqref{eq:tcmb}. Finally, we carried out a statistical analysis to test three models,
describing both spatial variations of $\Delta \alpha/\alpha$ and of $T_{CMB}$ itself. All models allow for monopole and
dipole amplitudes, and one of the models also includes a dependence on the look-back time. We used a MCMC algorithm to
explore different configurations  of the parameter space that are summarized in Table \ref{tab:priors}.
All results of our analysis are summarized in Table \ref{tab:results}, and are compared with the ones from other groups
in Table \ref{tab:anomalies}, Figures \ref{fig5} and \ref{fig6}.

At the present time clusters of galaxies are not competitive with high-resolution spectroscopic measurements in
absorption systems along the line-of-sight of bright quasars, but they nevertheless play an important role since they
offer the possibility to constrain such spatial variations in a totally different redshift range with respect to the one
tested by QSO, with different systematic vulnerabilities. Our analysis does improve by an order of magnitude the constraints
on spatial variations of $\alpha$ obtained by the Planck Collaboration \cite{planck_int_24} and by a factor $\sim2.5$
the analysis using a different galaxy cluster sample \cite{Galli2013}.

Finally, we note that the future is particularly promising. 
{ Not only has the number of X-ray selected, very massive clusters at redshifts ranging 
from $z{\sim}0.2$ to well beyond $z{=}0.7$ now exceeded a thousand \citep{ebeling01,ebeling13}, 
thus offering greatly improved statistics \textit{and} depth compared to the sample used by us here.}
With next-generation full sky CMB missions such as
COrE/PRISM \cite{PRISM} being able to identify several tens of thousands of SZ clusters and 
significantly extending the range of the redshifts where they are detected, this technique 
can ideally complement the spectroscopic measurements, with the latter (which individually 
are more precise) being fewer in number (say of order one hundred) and focusing on higher 
redshifts. More detailed analyses (including tomographic ones) as well as specific comparisons 
with particular classes of scalar field based models which include spatial and/or environmental
dependencies will then become possible. 

\section*{Acknowledgments}
We are grateful to Fernando Atrio-Barandela,  Gemma Luzzi and Ricardo G\'{e}nova-Santos for
useful conversations on the topic of this work. IDM acknowledges
the financial support from the University of the Basque Country UPV/EHU under the program
"Convocatoria de contrataci\'{o}n para la especializaci\'{o}n de personal 
investigador doctor en la UPV/EHU 2015", and from the Spanish Ministry of 
Economy and Competitiveness through research project FIS2014-57956-P (comprising FEDER funds).
This article is based upon work from COST Action CA1511 Cosmology and Astrophysics 
Network for Theoretical Advances and Training Actions (CANTATA), 
supported by COST (European Cooperation in Science and Technology).

It was also done in the context of project PTDC/FIS/111725/2009 (FCT, Portugal), 
with additional support from grant UID/FIS/04434/2013. CJM is also supported by an 
FCT Research Professorship, contract reference IF/00064/2012, funded by FCT/MCTES 
(Portugal) and POPH/FSE (EC). CJM thanks the Galileo Galilei Institute for Theoretical 
Physics for the hospitality and the INFN for partial support during the completion of this work.

\end{document}